# Crossover from two-dimensional to three-dimensional superconducting states in bismuth-based cuprate superconductor

Jing Guo[1]*, Yazhou Zhou[1]*, Cheng Huang[1,3]*, Shu Cai[1,3], Yutao Sheng[1,3], Genda Gu[2], Chongli Yang[1], Gongchang Lin[1,3], Ke Yang[4], Aiguo Li[4], Qi Wu[1], Tao Xiang[1,3,5] and Liling Sun[1,3,5]†

[1]*Institute of Physics, National Laboratory for Condensed Matter Physics, Chinese Academy of Sciences, Beijing 100190, China*
[2]*Condensed Matter Physics & Materials Science Department, Brookhaven National Laboratory, NY, 11973-5000 USA*
[3]*University of Chinese Academy of Sciences, Beijing 100190, China*
[4]*Shanghai Synchrotron Radiation Facilities, Shanghai 201204, China*
[5]*Songshan Lake Materials Laboratory, Dongguan, Guangdong 523808, China*

To decipher the mechanism of high temperature superconductivity, it is important to know how the superconducting pairing emerges from the unusual normal states of cuprate superconductors, including pseudogap, anomalous Fermi liquid and strange metal (SM). A long-standing issue under debate is how the superconducting pairing is formed and condensed in the SM phase because the superconducting transition temperature is the highest in this phase. Here, we report the first experimental observation of a pressure-induced crossover from two- to three-dimensional superconducting states in the optimally-doped $Bi_2Sr_2CaCu_2O_{8+\delta}$ bulk superconductor at a pressure above 2.8 GPa, through state-of-the-art *in-situ* high-pressure measurements of resistance, magnetoresistance and magnetic susceptibility. By analyzing the temperature dependence of resistance, we find that the two-dimensional (2D) superconducting transition exhibits a Berezinski-Kosterlitz-Thouless-like behavior. The emergence of this 2D superconducting transition provides direct and strong evidence that the SM state is predominantly 2D-like. This is important to a thorough understanding of the phase diagram of cuprate superconductors.

Since the discovery of superconductivity in the Ba-doped cuprate $La_2CuO_4$ in 1986[1], many breakthroughs in searching for new cuprates with higher superconducting transition temperature ($T_C$) have been achieved[2-8]. Up till now, more than two hundred cuprate superconductors, grouped into seven families with two fashions of hole and electron doping, have been found[2-10]. Structurally, the hole-doped cuprate superconductors hold peculiar octagonal or pyramid lattice with apical oxygen, which intrinsically leads to a complicated and unsteady lattice upon cooling due to the Jahn-Teller effect. Therefore, their normal states above the superconducting (SC) state, such as pseudogap (PG) [11,12], strange metal (SM) [13,14] and anomalous Fermi liquid (FL)[15,16], are full of the unknown physics of determining superconductivity[17,18]. Although a lot of theoretical progress on the superconducting mechanism of these high-$T_C$ superconductors has been made[19-27], a unified understanding on how the SC state connects with these unusual normal states is still lacking[28].

Because the SM state of the optimally-doped superconductor not only can develop the SC state with the highest $T_C$ but also links the PG and the anomalous FL states, it is of great interest to take the SM state as a breakthrough point to reveal the underlying physics of cuprate superconductors. In this high-pressure study, we chose the optimally-doped $Bi_2Sr_2CaCu_2O_{8+\delta}$ (Bi-2212) single crystal, a typical nearly-two-dimensional high-Tc superconductors with the SM normal state and widely studied in recent years [10,11,29-36], as the investigated material. We performed *in-situ* high-pressure measurements of resistance, magnetoresistance and alternating current (*ac*) susceptibly on the samples, with the attempt to reveal the connection between the SM state and the SC state.

Figure 1a shows the plots of temperature ($T$) versus in-plane resistance ($R_{ab}$) for one of the high quality Bi-2212 samples subjected to the pressures ranging from 0.97 GPa to 13.7 GPa. It is seen that the $R_{ab}(T)$ measured at 0.97 GPa displays a T-linear behavior over the temperature range above its onset $T_C$ (~96 K), manifesting that the sample is nearly in the ambient-pressure SM normal state and in an optimally-doped superconducting state [13,14]. Unexpectedly, at the pressure of ~ 2.8 GPa, we found a small resistance drop at the temperature about 20 K higher than its ambient-pressure $T_C$. This higher-temperature drop becomes more visible at ~ 5 GPa and pronounced on further compression. To characterize the higher-temperature resistance drop emerging from the SM state, we applied the magnetic field perpendicular to the *ab*-plane for the sample subjected to 9 GPa, and found that the drop is continuously suppressed by the magnetic field until it vanishes at ~1 T (Fig.1b). To confirm this pressure-induced "two-step drop behavior", we repeated the measurements with new samples for five independent runs, and proved that the results were reproducible. Figure 1c shows the results obtained from one of the experimental runs, and demonstrates that the higher-temperature resistance drop appears at ~ 2.9 GPa and it prevails up to 10.2 GPa, the maximum pressure of this run. Application of the perpendicular magnetic field on the compressed sample at 10.2 GPa, the similar behavior that the higher-temperature resistance drop shifts to the lower temperature upon increasing magnetic field and disappears at ~ 1T was observed (Fig.1d). These results indicate that the higher-temperature drop is associated with a superconducting transition.

To further identify the difference of these two superconducting states, we concurrently performed *in-situ* high-pressure measurements for the in-plane resistance ($R_{ab}$) and out-plane resistance ($R_c$) of our sample in the same setup. As shown in (Fig. 2a-2d), $R_c(T)$ of the sample subjected to the pressure from 6.0 GP to 9.0 GPa shows a non-metallic feature before the superconducting transition, while $R_{ab}(T)$ displays a metallic behavior with T-linear characteristic. Looking in detail on $R_c(T)$, we find a perturbation at ~116 K (Fig.3a) in the monotonically increased $R_c(T)$ upon cooling until an abrupt drop appears at ~90 K (6 GPa), where the lower-temperature resistance drop in $R_{ab}(T)$ presents coincidently. Therefore, the coincident resistance drops at 90 K signify the bulk (3D) superconducting transition in the sample. Significantly, when we plotted the curves of $R_c(T)$ and $R_{ab}(T)$ obtained at the same pressure in the same figure, it is found that the perturbation temperature in $R_c(T)$ is in excellent agreement with the temperature of the higher-temperature resistance drop in $R_{ab}(T)$ for each pressure point (Fig.2a-2d). Here, we define the higher-temperature $T_C$ as $T_C^{'}$ and the lower- temperature $T_C$ as $T_C^{3D}$ respectively.

To specify the observed behavior in further detail, we are the first to perform the combined *in-situ* high-pressure measurements of *ac* susceptibility and resistance for the same sample in the same diamond anvil cell. This demanding combined measurement for the studies of superconductivity under high pressure is full of technical challenges because the integration of the standard four-probes for the resistance measurement and the coils for the *ac* susceptibility measurements into the same pressure cell is very difficult. Figure 2e-2h show the plots of *ac* susceptibility ($\Delta\chi'$) versus temperature obtained at different pressures. The superconducting transition of the sample can be identified by the onset of deviation in the signal from the almost constant background on the high-temperature side (see the blue plots) and the plunge to zero resistance (see the red plots). At ~ 0.8 GPa, the zero resistance and the diamagnetism of the ambient-pressure superconducting phase with a 3D nature are clearly detected. However, the diamagnetic signal of the higher-temperature superconducting state observed in $R_{ab}(T)$ is failed to be captured from our *ac* susceptibility measurements throughout the pressure investigated. This indicates that the pressure-induced higher-temperature superconducting state in the optimally doped Bi-2212 may have an unusual nature.

To know the peculiarity of the higher-temperature superconductivity, we applied the magnetic field perpendicular ($B_\perp$) and parallel ($B_{//}$) to the *ab*-plane for the compressed sample at 10.1 GPa, respectively. As is seen in Fig.3a, Fig.S2 and Fig.S3 in Supplementary Information (SI), the superconductivity shows strongly anisotropic characteristics, *i.e.* sustainable up to 7 T under the parallel magnetic field while suppressed above ~ 0.5 T under the perpendicular magnetic field. We derived the scaling behavior of the $B_\perp$ and the $B_{//}$ as a function of temperature for the higher-temperature superconducting state and found a distinct temperature dependence of $B_\perp(T)$ and $B_{//}(T)$, namely $B_\perp(T) \propto (1 - T/T_C)$ versus $B_{//}(T) \propto (1 - T/T_C)^{0.5}$

(Fig.3b). The result of our scaling analysis leads to $B^2_{//}(T) \sim 700 \times B_{\perp}(T)$, a typical feature for a 2D superconductivity [37,38].

The remarkably anisotropic behavior of the higher-temperature superconductivity is reminiscent of the Berezinskii-Kosterlitz-Thouless (BKT) transition at which vortex-antivortex pairs unbind[39]. It is known that the BKT transition in 2D superconductors can be evidenced by the temperature ($T$) dependence of resistance ($R$) via $R_{ab}(T) = R_0 exp[-b(T/T_{BKT} - 1)^{-1/2}]$ (here, $R_0$ and $b$ are material parameters and $T_{BKT}$ is the BKT transition temperature)[39-41]. As illustrated in Fig.3c-3h, the obtained results from our $R_{ab}(T)$ measured at different pressures are consistent with the BKT-like behavior of the 2D film superconductors [37,38], suggesting that the higher-temperature superconductivity is a 2D superconductivity.

We summarize our experimental results in the pressure-$T_C$ phase diagram (Fig.4). It is seen that the 2D-SC state with a BKT-like behavior emerges from the SM state above a critical pressure ~2.8 GPa, and then the $T_C$ of the 2D-SC state displays the same trend as that of the 3D–SC state upon compression. Significantly, the emergence of this 2D superconducting transition provides direct and strong evidence that the SM state is predominantly 2D-like[42].

In addition, our high pressure X-ray diffraction measurements show that the sample holds its ambient-pressure structure up to the maximum pressure (~14.4 GPa) of this study (Fig.S4 in SI), implying that the crossover from 2D to 3D superconductivity is not related to the pressure-induced structural phase transition.

Unexpectedly, the same high-pressure measurements on the underdoped and overdoped Bi-2212 samples found no such a crossover (Fig.S5 in SI). These results suggest that the PG state in the underdoped superconductor and the anomalous FL state in the overdoped superconductor may prevent the formation of the pressure-induced 2D superconducting state or mask the resistance feature of this state.

It is noteworthy that the pressure-induced crossover from a 2D to a 3D superconducting states found in this study has not been reported in any other compressed bulk cuprate superconductors, despite it was observed in the $La_{2-x}(Sr/Ba)_xCuO_4$ cuprate under magnetic fields[43]. Therefore, this finding may provide a new clue that it may play a role as a breakthrough in achieving a better understanding on the high-$T_C$ superconductors. Consequently, some questions stimulated by these results are posed, such as whether the crossover is associated with the interaction between the Jahn-Teller effect and the high pressure effect? In addition, why can such a crossover be observed in the optimally doped Bi-2212 superconductor by pressure, but it cannot be captured in the underdoped or overdoped superconductors; and whether can such a phenomenon be observed in any other hole-doped cuprate superconductors. The answers for the questions may be of great interest for achieving a unified understanding on the cuprate superconductors, and deserve further investigations from the sophisticated theoretical studies and the challenged high-pressure measurements with the capacity of detecting dynamic physics.

**Acknowledgements**

We thank Qi-Kun Xue, Xu-Cun Ma, Jian Wang, Dung-Hai Lee and Hong Yao for very helpful discussions. This work in China was supported by the National Key Research and Development Program of China (Grant No. 2017YFA0302900, 2016YFA0300300 and 2017YFA0303103), the NSF of China (Grant Numbers 11427805, U1532267, and 11604376), the Strategic Priority Research Program (B) of the Chinese Academy of Sciences (Grant No. XDB25000000). J. G. is grateful for the support from the Youth Innovation Promotion Association of CAS (2019008). The work at Brookhaven National Laboratory was supported by the Office of Basic Energy Sciences, Division of Materials Sciences and Engineering, U.S. Department of Energy under Contract No. DE-SC0012704.

Correspondence and requests for materials should be addressed to L.S. (llsun@iphy.ac.cn).

* These authors contributed equally to this work.

(1979).

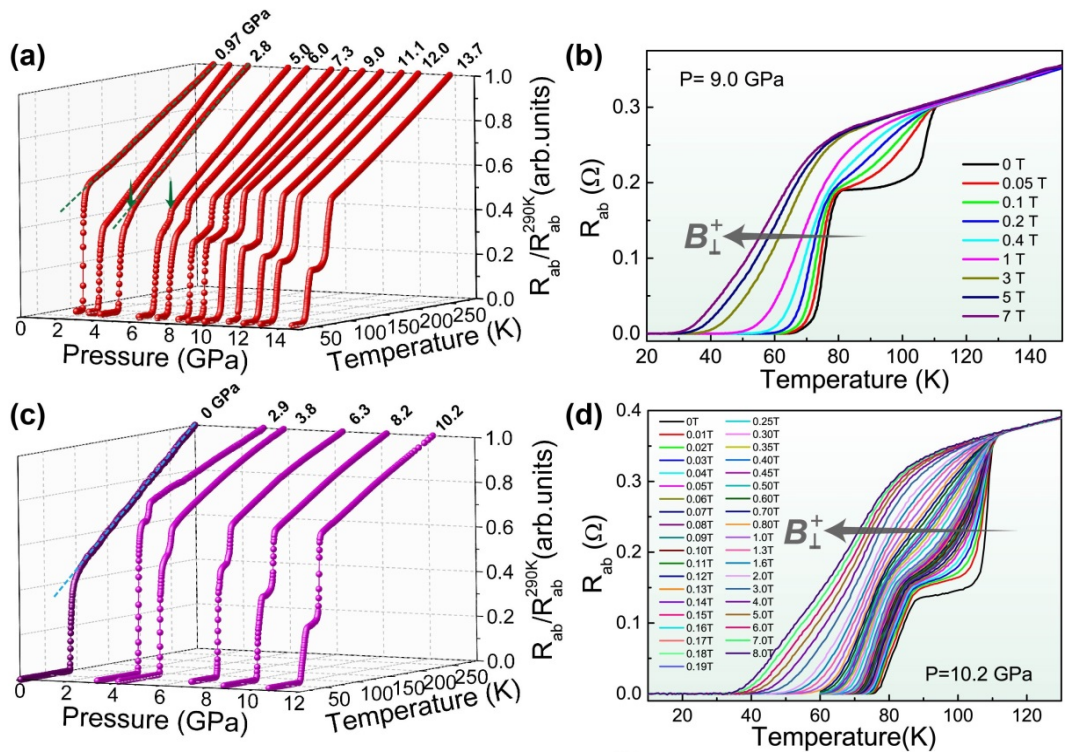

**Figure 1 The characterizations of the superconducting properties for the optimally doped Bi$_2$Sr$_2$CaCu$_2$O$_{8+\delta}$ superconductors under pressure.** (a) and (c) Temperature dependence of in-plane electrical resistance at different pressures for two experimental runs. (b) and (d) Magnetic field dependence of superconducting

transition temperatures measured on the sample subjected to 9.0 GPa and 10.2 GPa, respectively.

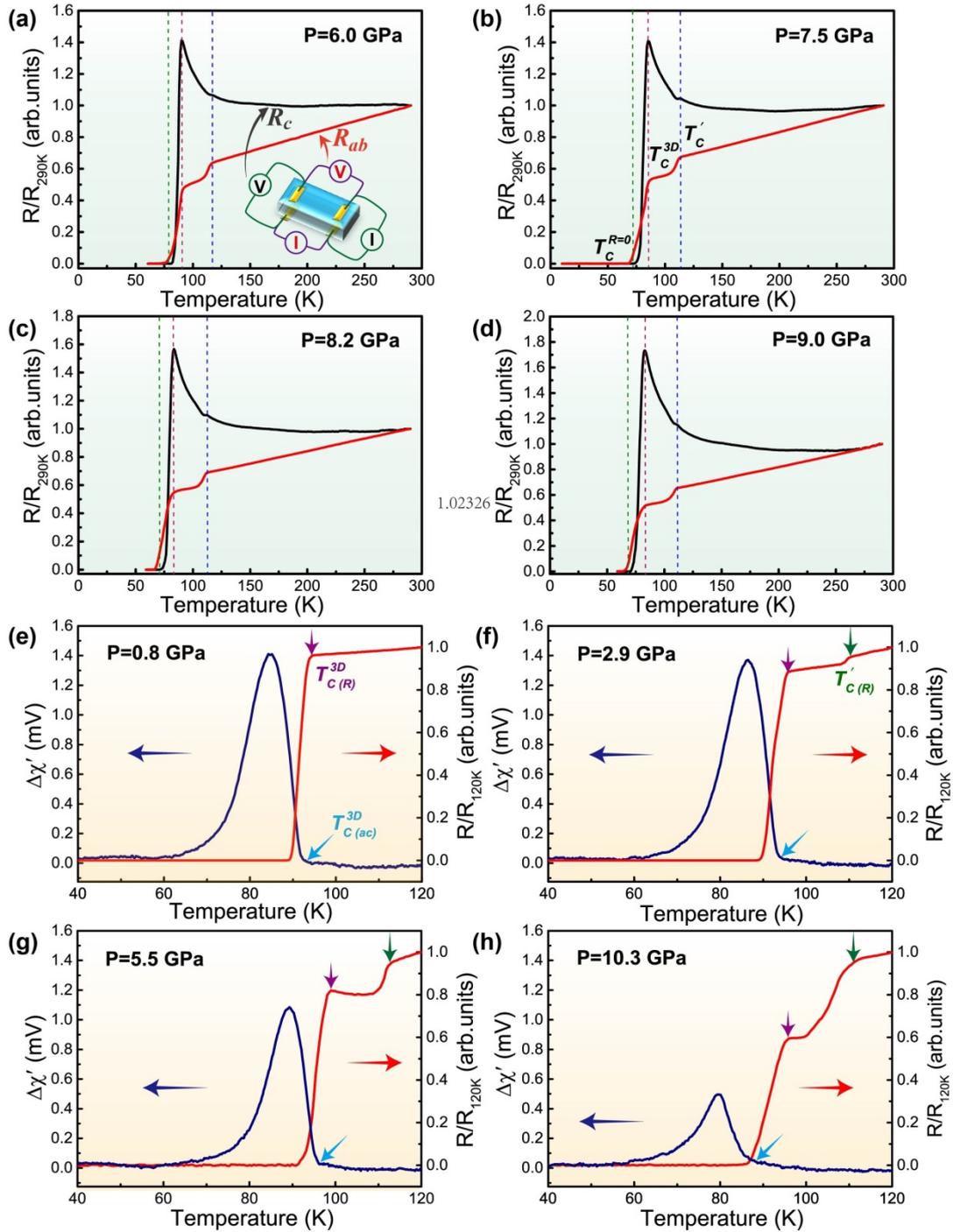

**Figure 2 The in-plan resistance ($R_{ab}$), out-plane resistance ($R_c$) and *ac* susceptibility ($\Delta\chi'$) as a function of temperature for the optimally-doped**

**Bi$_2$Sr$_2$CaCu$_2$O$_{8+\delta}$ superconductor.** (a)-(d) $R_{ab}(T)$ and $R_c(T)$ measured at different pressures. The inset of the figure (a) displays the integration of electrodes for the concurrent measurements on the $R_{ab}(T)$ and the $R_c(T)$. $T_C^{R=0}$ and $T_C^{3D}$ stand for the zero resistance and the onset of 3D superconducting transition temperatures, $T_C^{'}$ represents the onset temperature of the higher-temperature superconducting transition. (e)-(h) Temperature dependences of *ac* susceptibility ($\Delta\chi'$) and normalized electrical resistance ($R/R_{120K}$) at different pressures. The blue lines are the data of $\Delta\chi'(T)$, while the red lines are the data of $R/R_{120K}(T)$. The purple and green arrows indicate the temperatures of the 3D superconducting transition and the higher-temperature superconducting transition, respectively, shown in $R/R_{120K}(T)$. The cyan arrow indicates the temperature of the 3D superconducting transition probed by the *ac* susceptibility measurements.

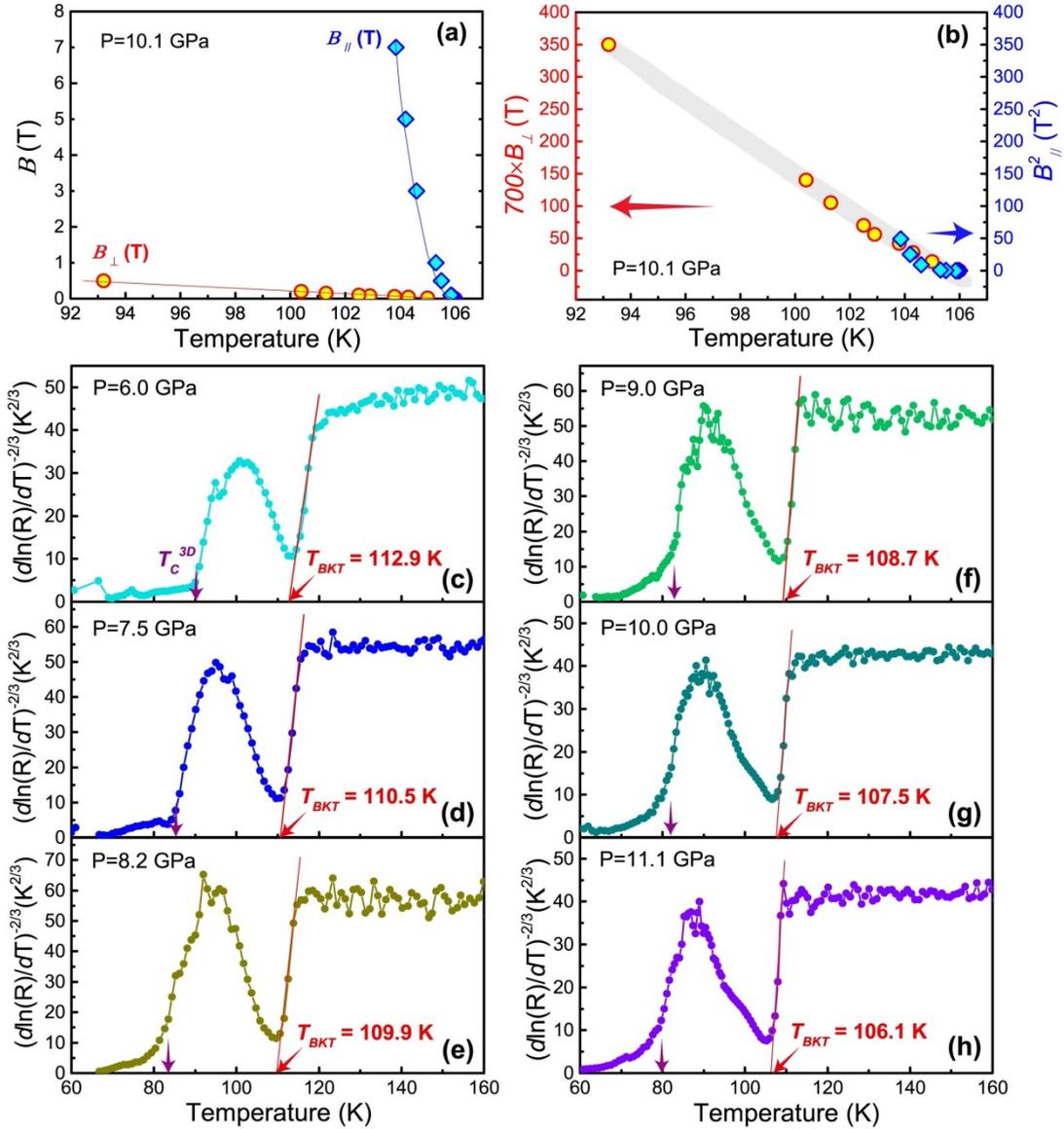

**Figure 3 Analyzing results of the 2D superconducting properties for the optimally-doped Bi$_2$Sr$_2$CaCu$_2$O$_{8+\delta}$ superconductors.** (a) Temperature dependence of the magnetic fields $B_\perp$ and $B_{//}$, extracted from Fig.S2 in Supplementary Information. (b) Scaling analysis for the magnetic fields $B_\perp$ and $B_{//}$ versus temperature ($T$), yielding that $B_{//}^2$ is about 700 times higher than $B_\perp$. (c)-(h) The plots of $dln(R_{ab})/(dT)^{-2/3}$ versus temperature, derived from the form of $R_{ab}(T) = R_0 exp[-b(T/T_{BKT} - 1)^{-1/2}]$, at different pressures. The purple and red arrows indicate the temperatures of the 3D

superconducting transition and the BKT-like transition, respectively.

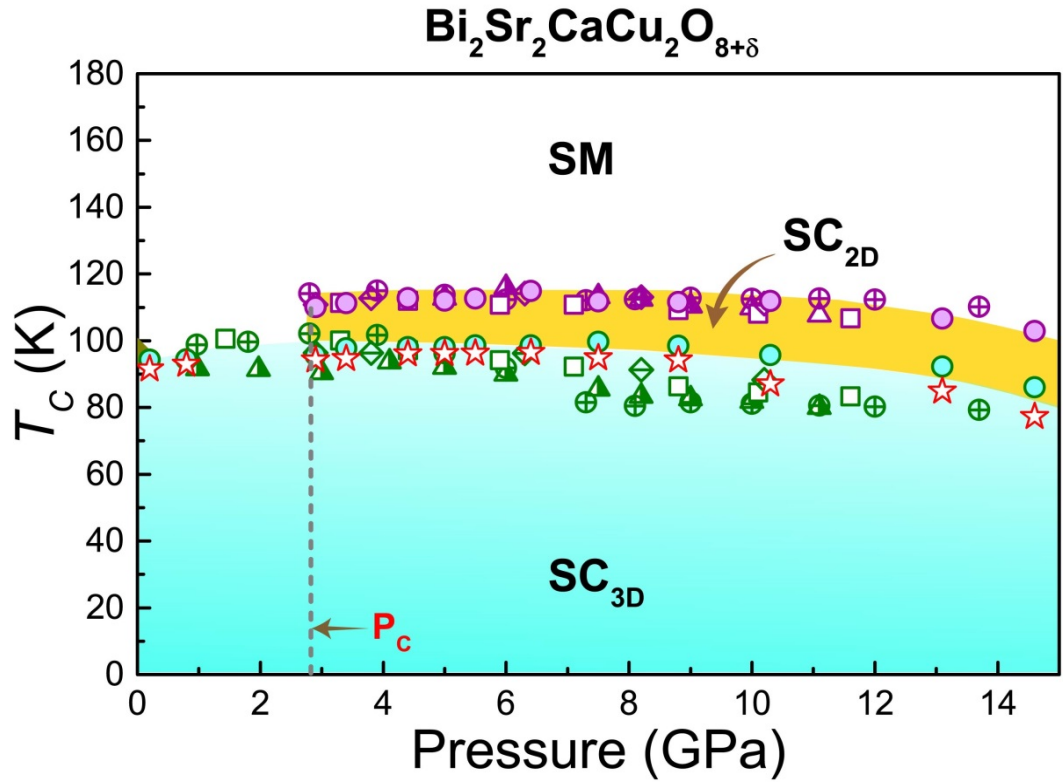

**Figure 4 Pressure-$T_C$ phase diagram established by the results obtained from different experimental runs for optimally-doped $Bi_2Sr_2CaCu_2O_{8+\delta}$ superconductor.** The acronyms of $SC_{2D}$ and $SC_{3D}$ stand for 2D (BKT-like) and 3D superconducting states, respectively. SM represents strange metal state. $P_C$ represents the critical pressure above which the 2D superconductivity emerges from the SM state.